\documentclass[journal=jacsat,manuscript=article]{achemso}

\usepackage{atbegshi}
\AtBeginDocument{\AtBeginShipoutNext{\AtBeginShipoutDiscard}}

\usepackage[T1]{fontenc} 
\usepackage{mathrsfs}    

\usepackage{amssymb,array}
\usepackage[superscript]{cite}
\usepackage{url}
\usepackage[utf8]{inputenc}
\usepackage[T1]{fontenc}
\usepackage{color}
\usepackage{epstopdf}
\usepackage{ae}
\usepackage{amsmath}
\usepackage{amsfonts}
\usepackage{graphics}
\usepackage{graphicx}
\usepackage{multirow}
\usepackage{multicol} 
\usepackage{bm}
\usepackage{float}
\usepackage{adjustbox}
\usepackage{colortbl} 
\usepackage{xcolor} 
\usepackage[most]{tcolorbox}
\usepackage{array, multirow, multicol, tabularx}
\usepackage{ulem}  

\definecolor{lightpurple}{rgb}{196,77,255}
\definecolor{lightyellow}{rgb}{255,233,0.8}
\definecolor{lightblue}{rgb}{0.2,0.6,1}
\definecolor{lightred}{rgb}{1,0.5,0.5}
\definecolor{green}{rgb}{0.5,100,0}

\usepackage{tikz}
\usepackage{collcell}
\usepackage{xstring}

\author{Andy D. Zapata-Escobar}
\affiliation[unne]{Institute for modeling and innovative technology, IMIT (CONICET-UNNE), Corrientes, Argentina}
\alsoaffiliation[imit]{Institute for modeling and innovative technology, IMIT (CONICET-UNNE), Corrientes, Argentina}
\author{Srimanta Pakhira}
\affiliation{Department of Physics, Department of Metallurgy Engineering and Materials Science (MEMS), Centre for Advanced Electronics (CAE), Indian Institute of Technology Indore (IIT Indore), Simrol, Khandwa Road, Indore, 453552, Madhya Pradesh (M.P.), India.}
\author{Joaquin Barroso-Flores}
\affiliation{Instituto de Química, Universidad Nacional Autónoma de México (UNAM), Circuito Exterior SN, Ciudad Universitaria, Coyoacán CP 04510 Ciudad de México, México}
\author{Gustavo A. Aucar}
\affiliation{Physics Department, Natural and Exact Science Faculty, Northeastern University of Argentina}
\alsoaffiliation[imit]{Institute for modeling and innovative technology, IMIT (CONICET-UNNE), Corrientes, Argentina}
\email{gaaucar@conicet.gov.ar (G.A.A.)}
\author{Jose L. Mendoza-Cortes}
\affiliation{Department of Physics, Scientific Computing, Material Sciences and Engineering, High-Performance Material Institute, Condensed Matter - High Magnetic Field National Lab, Florida State University, Tallahassee FL, 32310, USA.}
\alsoaffiliation{Department of Chemical and Biomedical Engineering, FAMU-FSU Joint College of Engineering, Tallahassee, FL 32310, USA.}
\alsoaffiliation{Current address: Department of Chemical Engineering \& Materials Science, Michigan State University, East Lansing, Michigan 48824, USA.}
\email{jmendoza@msu.edu  (J.L.M-C.)}

\title{Relativistic Quantum Calculations to Understand the Contribution of f-type Atomic Orbitals and Chemical Bonding of Actinides with Organic Ligands}

\keywords{Magnetic shielding \sep LRESC \sep Relativistic effects \sep ...}

\begin{document}
\begin{abstract}
The nuclear waste problem is one of the main interests of the rare earth and actinide elements chemistry. Studies of Actinide-containing compounds are at the frontier of the applications of current theoretical methods due to the need to consider relativistic effects and approximations to the Dirac equation in them. Here, we employ four-component relativistic quantum calculations and scalar approximations to understand the contribution of f-type atomic orbitals in the chemical bonding of actinides (Ac) to organic ligands. We studied the relativistic quantum structure of an isostructural family made of Plutonium (Pu), Americium (Am), Californium (Cf), and Berkelium (Bk) atoms with the redox-active model ligand; DOPO (2,4,6,8-tetra-tert-butyl-1-oxo-1H-phenoxazin-9-olate). Crystallographic structures were available to validate our calculations for all mentioned elements except for Cf. In short, state-of-the-art relativistic calculations were performed at different levels of theory to investigate the relativistic effects and electron correlations on geometrical structures and bonding energies of $Ac$-DOPO$_3$ complexes ($Ac$=Pu, Am, Cf, Bk) : 1) the scalar relativistic zeroth order regular approximation (ZORA) within the hybrid density functional theory (DFT) and 2) the four-component Dirac equation with the Dirac-Hartree-Fock (4c-DHF) and Lévy-Leblond (LL) Hamiltonians. We show that scalar DFT-ZORA could be used as an efficient theoretical approximation to first approximate the geometry and electronic properties of actinides which are difficult to synthesize or characterize; but knowing that the higher levels of theory, like the 4c-DHF, gives closer results to experiments than the scalar DFT-ZORA. We also performed spin-free calculations of geometric parameters for the Americium and Berkelium compounds. 
To the best of our knowledge, this is the first time that such kind of large actinide compounds (the largest contains 67 atoms and 421 electrons) are studied with highly accurate four-component methods (all-electron calculations with 6131 basis functions for the largest compound). We show that relativistic effects play a key role in the contribution of f-type atomic orbitals to the frontier orbitals of $Ac$-DOPO$_3$ complexes. The analysis of results obtained applying different theoretical schemes to calculate bonding energies are also given.
\end{abstract}

\section{Introduction} \label{sec:intro} 
%
Our understanding of how chemistry and physical properties evolve across rows and down groups in the modern periodic table has been at the forefront of chemistry for the last 150 years. The latest additions the so-called actinide elements, which are the heaviest elements, are at the frontier of our current chemical and physical understanding.\citep{Ohtsuka2019,greenwood1997chemistry,galley2019synthesis,brenner2020trivalent,leszczyk2019new} The radioactive actinide elements residing underneath the main body of the periodic table conjure images of mushroom clouds, glowing test tubes, and superheroes awakening. While there is some truth to this view, as chemists working beyond the edge of nuclear stability, we have peered into the abyss to find elements whose realities are so complex and mystifying that we do not need to invent fanciful tales to remain a captive audience. Our fundamental knowledge and understanding of actinide bonding trends and coordination modes is significantly limited compared to other elemental series in the periodic table. Even though there are several and thorough studies involving actinide elements,6-10 there are no studies that use state-of-the-art four-component relativistic methods that permits to grasp on the contribution of f-type atomic orbitals (AOs) in the electronic structures and chemical bonding of actinides (Ac) to medium-size organic ligands. This is needed to understand the changes in interactions and chemical bonds, which may impact speciation and separation selectivity to different heavy elements.\citep{Ohtsuka2019,greenwood1997chemistry,galley2019synthesis,brenner2020trivalent}

Some chemical reactions are clearly explained by considering relativistic effects on valence electrons of heavy-atoms, e.g. the electromotive force of the cells of lead-batteries, where the relativistic effects on Pb atoms allows for the battery to work.\citep{leszczyk2019new} There are also few other examples which are explained by relativistic effects: e.g., the yellow color of gold, the yellow color of hexachloroplumbate(IV), the catalytic methane activation by Pt$^+$, to name a few.\citep{pyykko2012relativistic} From these grounds, we have undertaken a research program whose goals are oriented to learn more about the physics that underlies the structure and electronic properties of actinides bonding, where one must consider relativity. Here, we present our first results of this research program, by studying actinide containing compounds having Pu, Am, Cf and Bk and an organic ligand model. Accordingly, we show how the 5f electrons of these actinides behaves differently when relativity is taken into account, thus changing the frontier molecular orbitals of complexes. More specifically, we estimate relativistic effects at different levels of theory:  ZORA--HF, ZORA--DFT, ZORA--DFT-D3, and LL and DHF. More specifically, a) the spin-orbit (so) in addition to the scalar (sr) relativistic zeroth order regular approximation in hybrid density functional theory (ZORA--DFT)\citep{chang1986regular, van1994relativistic} and without DFT (ZORA-HF), and b) four-component (4c) Dirac equation with the Dirac-Hartree-Fock (DHF), Lévy-Leblond (LL) and spin-free (SF) Hamiltonians.\citep{saue2011relativistic} In this manner, it was possible to learn about the influence of both, relativistic and electron correlation effects on the geometry, the energy involved in the bonding process and the change of 5f and 6d AOs in the population of molecular orbitals. To the best of our knowledge, this is the first time that such kind of large actinide compounds are studied with highly accurate four-component methods.

The research for carbon-free energy has generated a tremendous interest in alternative energy sources. Nuclear energy is a promising alternative because it offers opportunities for low cost and high efficiency. However, it generates a significant amount of radioactive waste, which we are not able to process or reduce manly because of our lack of understanding of bonding in heavy elements. The proper disposal of nuclear waste is still challenging and one of the most difficult kinds of waste to handle because it is highly hazardous and harmful for humans and the biosphere even for small concentrations, and especially if they are radioactive. Advanced actinide separation strategies reduce hazards and costs of managing radioactive waste. Among them liquid fluoride thorium reactor and other molten salt reactor (MSR) have been designed to use the thorium fuel cycle with a fluoride-based, molten, liquid salt for fuel. These reactors are passively safe, produce less radioactive waste, offer higher fuel burn up efficiency, and do not produce industrial quantities of fissile $^{239}$Pu.\citep{Ohtsuka2019} Similarly, the development of vitrification using borate chemistry was inspired, in part, from evidence that a significant boron concentration exists in the brine solutions found at the Waste Isolation Pilot Plant (WIPP) in Carlsbad, New Mexico.\citep{goel2019,vienna2010nuclear}

Actinide elements are of considerable technological importance. In condensed matter, as they may be high temperature superconductors or centers of lasing activity.\citep{galley2019synthesis,fisk1988heavy} Recently, various studies have been performed in lanthanides and actinide elements to understand the f-orbitals expansion and contractions when loosing or gaining electrons.\citep{galley2019synthesis,brenner2020trivalent,van1999origin,boring2000plutonium} Therefore, the further understanding of the chemical similarities and differences between the trivalent lanthanides and the actinides analogues is of immense interest. One indication of these advances is the rapid increase in f-block element containing compounds for which high-resolution structures exist, while their properties have been investigated both experimentally and theoretically during the last decade.\citep{greenwood1997chemistry,galley2019synthesis} This is promising because the f-electrons remain quite localized in going from the atomic to the condensed state, thus, a lot of knowledge gained from atoms might be transferable to the condensed state.\citep{galley2019synthesis,van1999origin,boring2000plutonium,cao2021understanding} Therefore, understanding f-block elements and trends is essential, yet challenging because of the intrinsic radioactivity of most of the actinide elements.\citep{brenner2020trivalent,cooper2000challenges} However, this lack of knowledge poses challenges that prevents advances in nuclear energy, radioactive waste management, separation, storage, and next-generation f-electrons materials.\citep{cooper2000challenges}

\medskip

There has been an increasing interest in fundamental research about actinide elements since the 20th century. Understanding the chemical bonding, chemical reactivity, interaction, and speciation of hazardous heavy actinide elements in aqueous solutions are a first step toward modeling their behavior, both in the laboratory and under natural conditions. Although the chemistry of the actinides has been extensively explored for over a century, its theoretical grounds and basic understandings still lags much compared to that of most elements in the periodic table.\citep{galley2020evidence} The ability of actinides to form chemical bonds and the role of 5f and 6d AOs in the contribution to molecular orbitals (MO) are still subjects of investigation.\citep{tobisu2012rhodium,kundu2009rational} We are still developing our understanding of the role of 5f- and 6d-shell electrons from actinides elements when interacting with other atoms or molecules.

Recently, Silver et al. experimentally characterized and synthesized the berkelium coordination complex, Bk(III)tris(dipicolinate), and showed that it had formed a chemically distinct Bk(III) borate material.\citep{silver2016characterization} A recent study has synthesized nonaqueous isostructural family of f-element compounds (Ce, Nd, Sm, Gd; Am, Bk, Cf) of the redox-active dioxophenoxazine ligand (DOPO = 2,4,6,8-tetra-tert-butyl-1-oxo-1H-phenoxazin-9-olate) for the first time.\citep{galley2020evidence} Their electronic structure was further studied by applying the zeroth order regular approximation (ZORA) with hybrid DFT as well as CASSCF calculations to predict and investigate the structure, oxidation states, and electronic properties of both the lanthanides and actinides complexes in detail.\citep{galley2020evidence} More recently, Albrecht-Schmitt et al. showed that the complexation of berkelium(III) by carbonate results in spontaneous oxidation to berkelium(IV) and that multiple species can be present in solution.\citep{albrecht2020theoretical} They explored the chemistry of Bk(IV) carbonate and carbonate-hydroxide complexes based on theoretical comparisons with spectroscopic data previously reported for Bk(IV) carbonate solutions.\citep{albrecht2020theoretical} They performed molecular orbital calculations to try to obtain a better understanding of the nature of their chemical bonding using DFT at the level B3PW91. Brenner et al. synthesized and analyzed the structure, and solid-state UV--vis--NIR spectroscopy of three new f-element squarates, M$_2$(C$_4$O$_4$)$_3$(H$_2$O)$_4$, where M = Eu, Am, Cf and they showed that Cf has a 3$^+$ ionic state with the nine--coordinate ionic radius of Cf$^{3+}$ about 1.127 $\pm$ 0.003 \AA.\citep{brenner2020trivalent} Galley et al. synthesized the Cf(DOPO)$_2$(pyridine)(NO$_3$) complex and they carried out a computational analysis of the electronic structure of the complex to investigate the ligand-Cf$^{III}$ interactions.\citep{galley2020evidence} Goodwin et al. experimentally synthesized Am$^{3+}$ organometallic complex [Am(C$_5$Me$_4$H)$_3$] and applied DFT and ab initio wave function theory calculations to explore the interaction and electronic structure of the complex.\citep{goodwin2019back}

\medskip

In the next sections we shall first explain the methodology which we have implemented to calculate the geometry and bonding energies of the molecular compounds. We have considered two different levels of theory: 1) ZORA\citep{becke1993becke, lee1988development, grimme2006semiempirical, lenthe1993relativistic, van1996relativistic, van1996zero} and 2) 4c-DHF and LL, with different schemes of basis sets.\citep{lenthe1993relativistic, van1996relativistic, van1996zero} The main idea is to show how important relativistic effects are when one is interested to reproduce those properties. Afterwards we shall focus on one of the main aims of this work, namely, to learn about the dependence of the participation of p-, d- and f- AOs of the actinides atoms on the frontier electronic structure of the just mentioned family of compounds: $Ac$-DOPO ($Ac$=Pu, Am, Cf, Bk) with the above mentioned level of theory. We shall show that relativistic effects together with electron correlation modify in a great manner the contribution of those AOs, as well as the values of bonding energies. 
%
\section{Methods and Computational details} \label{sec:compdet}
%
All geometrical structures of the Pu-DOPO, Am-DOPO, Cf-DOPO, and Bk-DOPO complexes reported in the present study were fully optimized with different levels of theory: i) ZORA with hybrid DFT including dispersion corrections, i.e., ZORA-B3LYP-D3 or DFT for short,\citep{tobisu2012rhodium, kundu2009rational, silver2016characterization, albrecht2020theoretical, goodwin2019back, becke1993becke} ii) ZORA-HF, which means ZORA without electron correlation, calculated for only the Am-DOPO and Bk-DOPO compounds, and iii) 4c DHF/LL Hamiltonians with basis sets that are large enough for describing heavy atoms and those in the first coordination sphere to the heavy atoms.\citep{cremer2014dirac} We must mention here that we had performed 4c calculations to obtain geometrical structures at DFT level of approach but in this case we were unable to obtain converged results.

The first method essentially improves the usual non-relativistic hybrid DFT\citep{becke1993becke,lee1988development} methods and is useful for general chemistry applications. Long-range van der Waals (vdW) dispersion correction were included.\citep{grimme2006semiempirical}  The Grimme's--D3 terms were used for all the light atoms, but due to availability, the dispersion terms from the universal force field (UFF) were employed for actinides. The Slater--Type Orbital (STO) basis sets were used instead of Gaussian-type orbital (GTO) basis sets for the ZORA calculations. Valence triple-$\zeta$ with polarization functions (TZP) quality were used for Pu, Am, Cf, and Bk atoms, and the valence double-$\zeta$ with polarization functions (DZP) quality STO basis sets were used for the rest of the atoms (C, N, O, and H).\citep{van2003optimized} In general, the STO basis sets give relatively consistent and rapidly converging results.\citep{guell2008importance} 
We have also made ZORA calculations without any DFT functionals.

\medskip

The second procedure is grounded on the 4c-DHF Hamiltonian, and then its Lévy-Leblond model, to get appropriate non-relativistic (NR) results. In order to compare results obtained with different relativistic methods, though not including in any of them the spin-dependent contributions, we also performed SF calculations to compare them with those of sr-ZORA. All studies were performed by applying a local dense basis set scheme (LDBS),\citep{dyall2007relativistic} meaning basis sets that are as large as possible for heavy atoms, and then, less complete basis set for atoms that are not much involved in the main fragment of interest (second coordination sphere and beyond). To accomplish highly accurate four-component calculations we have designed an appropriate scheme of basis sets. For atoms different to carbon and hydrogen the selected basis set is dyall-cv3z,\citep{dyall2012core} while carbon and hydrogen atoms are treated with 3-21G Gaussian basis sets.\citep{binkley1980self,gordon1982self} Geometries of the complexed systems and that of each part (actinide atoms and DOPO molecules) were optimized using the DIRAC code.\citep{gomesdirac19} Some LDBS schemes were used to describe more accurately the influence of d-shell and f-shell AOs on the bonding of actinide atoms to the oxygen atoms and nitrogen atoms. Only neutral systems (i.e., Am and Bk), were considered at the DHF  and LL levels of theory because it is not feasible to do the geometry optimization of open-shell systems with analytical gradient in the current implementation of the DIRAC code. 

On the other hand, to calculate the bonding energies (BEs) at DHF and LL levels, we first considered the closed--shell systems though ionized, meaning Ac$^{+3}$ and DOPO$^{-1}$, and optimized their geometries. We did it by including at the begining of the calculation a higher density of charges on the nitrogen atoms (-1 each) and a lower charge density on the $Ac$ atoms (+3). Then, we calculated the energy of the DOPO structure (which is an open--shell system) by a single-point calculation. The energies of the actinides were also obtained from a single--point calculation with the DIRAC code at the DHF and LL levels of theory. On the other hand, the BEs of both actinides systems, Am--DOPO and Bk--DOPO were calculated at the ZORA level of approach without any difficulty. In all cases we have chosen to consider both, the close-shell configurations and the open-shell configurations as shown in the supplementary information (SI).

\begin{figure}[h!]
 \centering{
    \includegraphics[scale=0.6]{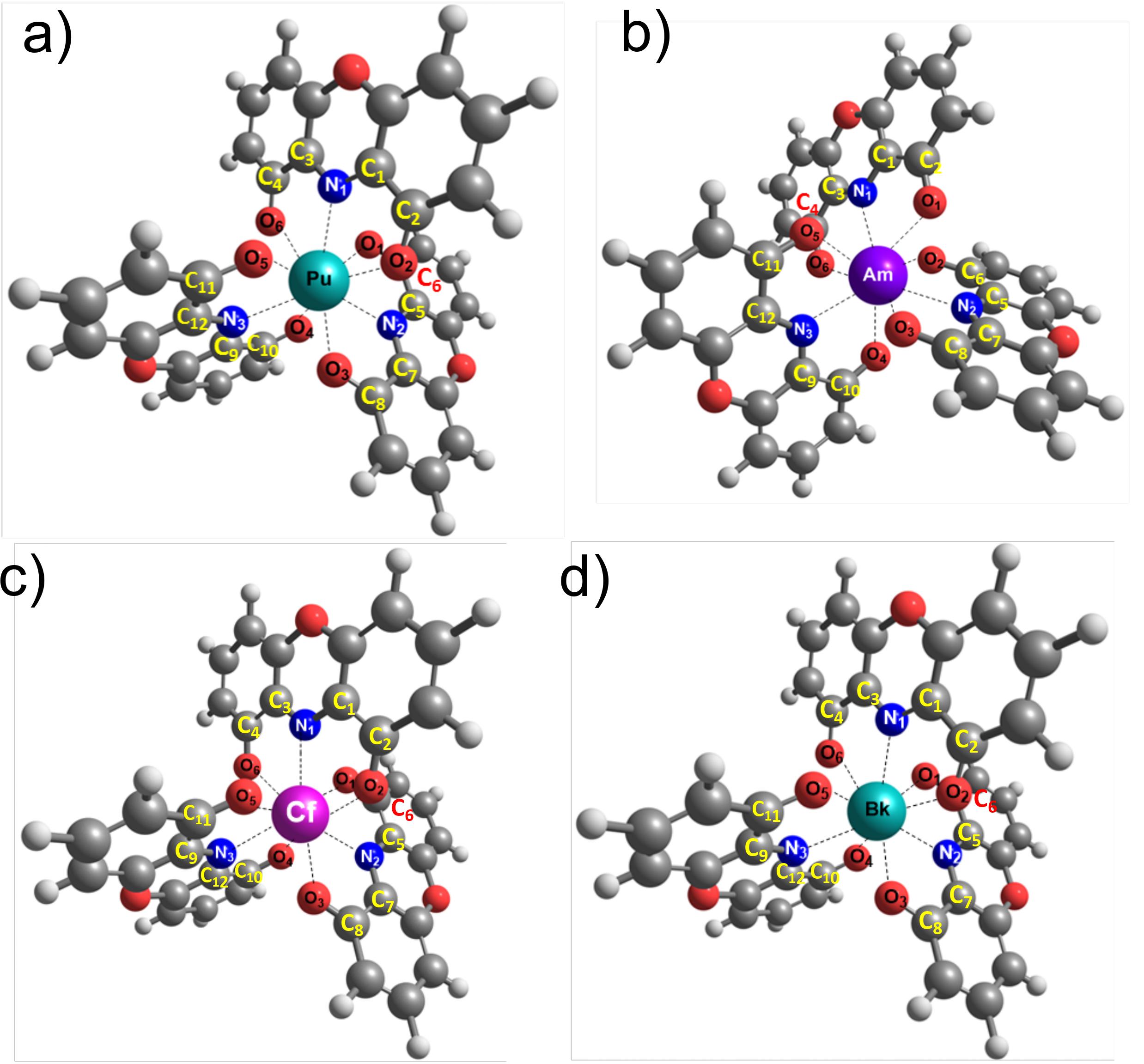}
 }
 \caption{Optimized structures of $Ac$-DOPO compounds with a) $Ac$=Pu, b) $Ac$=Am, c) $Ac$=Bk, and d) $Ac$=Cf.}
 \label{f:1}
\end{figure}

The whole set of model compounds considered in this work has been experimentally synthesized and characterized, and are denoted as $Ac$--DOPO (Figure \ref{f:1}). In all computations no constraints were imposed on the geometry. The scalar relativistic ZORA-DFT calculations in addition to the so relativistic ZORA-HF calculations were performed with the Amsterdam density functional (ADF) code\citep{te2001chemistry,ADF2001} while the 4c-DHF, 4c-LL and 4c-DHF-SF calculations were performed with the Dirac code.\citep{gomesdirac19} The ADF-GUI\citep{ADF2001} program was used for the visualization of the highest occupied molecular orbitals (HOMO) and lowest unoccupied molecular orbitals (LUMO) from the ZORA-DFT calculations. The ligand field diagram (LFD) analysis of the $Ac$-DOPO system ($Ac$ = Pu, Am, Cf and Bk) for all actinides was performed using the ZORA approximation, while DHF and LL levels of theory were used only for Am and Bk.

%

%
\section{Results and Discussion} \label{sec:results}
%
We shall first show the influence of relativistic effects on the geometric parameters of the actinide compounds, and then on BEs. Afterwards, we shall show how important are the contributions of each of the valence-shell p-, d- and f- AOs of actinides in the electronic framework of the frontier MOs, HOMO-LUMO.
%
\subsection{Geometrical structures}
%
The optimized geometry of Pu--DOPO complex at the ZORA level is shown in Figure \ref{f:1}a. The root mean square deviation (RMSD) value of all the 1$^{st}$ coordination bonds (Pu--N, Pu--O) w.r.t. to experiments is 0.104 \AA, while the RMSD for the 2$^{nd}$ coordination sphere (C--C, C--N, and C--O) is 0.046 \AA, as shown in Tables S1-S3. In other words, the calculations at ZORA--DFT--D3 level of approach estimate the experimental Pu--N and Pu--O bonds to 0.1 \AA of error, which is the order of magnitude for high quality x--ray structures.

\medskip

The optimized geometry of the Am--DOPO complex is shown in Figure \ref{f:1}b. In this case, we used four different levels of theory:  ZORA--HF, ZORA--DFT, ZORA--DFT--D3, 4c--LL and 4c--DHF (see Table \ref{t:1} and Table S4). The RMSD for the bond lengths of Am--N and Am--O are 0.206, 0.024, 0.052, 0.023, 0.050 and 0.149, 0.120, 0.106, 0.103, 0.066 \AA, respectively. This trend is in agreement with the increasing accuracy of the level of theory, being 4c-DHF the highest and the smallest values of RMSD is found for d(Am--O) though there are very small values of RMSD for d(Am--N). In this last case only at the ZORA--HF level we obtain RMSD values that are larger than 0.1 \AA. Results of 4c-DHF-SF calculations gives the same bond lengths as those of ZORA-DFT. Looking more closely at the bonds, the calculations with ZORA--DFT--D3 are within a difference of 0.105 \AA for Am-O and 0.046 \AA for Am--N (Table S5). When they are calculated with the Lévy--Leblond model, those differences are 0.102 \AA and 0.029 \AA, respectively. In the case of DHF calculations such differences are within 0.066 \AA and 0.054 \AA, respectively. This means that both, electron correlation and relativistic effects must be included to best reproduce experimental measurements, though relativistic effects showed by DHF calculations give the best reproduction of d(Am--N) and d(Am--O). Remarkably, ZORA with dispersion corrections improves the RMSD only a bit versus calculations without it.

\begin{table}[htbp]
\caption{The equilibrium bond distances of Am-N and Am-O at different levels of theory. All values are given in \AA.}

\adjustbox{max width=\textwidth}{%
\begin{tabular}{|c|cccccc|ccccc|}\hline
&\multicolumn{6}{|c}{Bond length } & \multicolumn{5}{|c|}{Difference w.r.t. experiments } \\ \hline
Bonds & Expt.$^{(a)}$ & ZORA-HF & ZORA-DFT & ZORA-DFT-D3 & 4c-LL & 4c-DHF
      & ZORA-HF & ZORA-DFT & ZORA-DFT-D3 & 4c-LL & 4c-DHF \\ \hline
Am--N$_1$&2.591 &  2.372 & 2.554 & 2.542 & 2.614 & 2.626 & -0.219 & -0.037 & 0.049  &-0.023&-0.035 \\
Am--N$_2$&2.582 &  2.369 & 2.563 & 2.522 & 2.601 & 2.621 & -0.213 & -0.019 & 0.06   & 0.019&0.039 \\
Am--N$_3$&2.554 &  2.369 & 2.546 & 2.507 & 2.581 & 2.623 & -0.185 & -0.008 & -0.047 & 0.027&0.069 \\
RMSD     &      &        &       &       &       &       & 0.206  & 0.024  & 0.052  & 0.023&0.05 \\
         &      &        &       &       &       &       &        &        &        &      & \\
Am--O$_1$&2.472 & 2.334 & 2.592 & 2.595 & 2.588 & 2.541 & -0.138 & 0.12 & 0.123 & 0.116 &0.069 \\
Am--O$_2$&2.465 & 2.321 & 2.614 & 2.573 & 2.571 & 2.536 & -0.144 & 0.149 & 0.108 & 0.106&0.071 \\
Am--O$_3$&2.476 & 2.314 & 2.557 & 2.558 & 2.556 & 2.538 & -0.162 & 0.081 & 0.082 & 0.08 &0.062 \\
Am--O$_4$&2.462 & 2.333 & 2.579 & 2.579 & 2.58 & 2.545 & -0.129 & 0.117 & 0.117 & 0.118 &0.083 \\
Am--O$_5$&2.481 & 2.324 & 2.621 & 2.586 & 2.581 & 2.539 & -0.157 & 0.14 & 0.105 & 0.1   &0.058 \\
Am--O$_6$&2.493 & 2.332 & 2.592 & 2.586 & 2.583 & 2.544 & -0.161 & 0.099& 0.093 & 0.09  &0.051 \\ \hline
RMSD     &      &       &       &       &       &       & 0.149  & 0.12 & 0.106 & 0.103 &0.066 \\ \hline
\end{tabular}}
\label{t:1}
$^{(a)}$ Taken from Ref. \citep{galley2019synthesis}.
\end{table}

All this analysis suggest that dispersion corrections might not be necessary to consider for this type of compounds, as they were not included in the 4c-LL and 4c-DHF calculations either. Considering that electron correlation effects can be obtained as the difference between ZORA--DFT and ZORA--HF calculations, we found that these effects increase both distances: in average 0.166 \AA for d(Am--N) and 0.267 \AA for d(Am--O). On the other hand, considering that relativistic effects can be obtained as the difference between DHF and LL calculations we found that they increase the values of d(Am--N) (0.024 \AA in average) but diminish that of d(Am--O) (0.037 \AA in average) though in a much smaller amount as compared with the influence of those distances by the electron correlation effects.

\medskip

The geometry of the Bk--DOPO complex was calculated with the three levels of theory mentioned above. In Figure \ref{f:1}c we show its equilibrium geometry. The bond distances Bk--N and Bk--O at equilibrium and for all methods used here are collected in Table \ref{t:2}. The RMSD is 0.082, 0.114 and 0.063 for ZORA--DF--D3, 4c-LL and 4c-DHF, respectively. The distances at 4c-DHF level of theory gives again the smallest value of RMSD, which one might expect since it is the most accurate. Similarly to what happens with $Ac$=Am, the bond lengths with 4c-DHF have the same consistent sign, though calculations at ZORA and 4c-LL levels underestimate the Bk--N bonds. Once both effects, electron correlation and relativistic effects are introduced, the behavior of those bond distances follows a similar trend as observed for Am. In this case, electron correlation effects are quite similar to those of relativistic effects for d(Bk--N).

\begin{table}[htbp]
\caption{The equilibrium bond distances Bk--N and Bk--O in the Bk--DOPO complex. All values are given in \AA.}
\adjustbox{max width=\textwidth}{%
\begin{tabular}{|c|ccccc|cccc|} \hline
&\multicolumn{5}{|c|}{Bond length } & \multicolumn{4}{c|}{Difference w.r.t. experiments }\\ \hline
Bonds&Expt.$^{(a)}$ & ZORA-HF & ZORA-DFT-D3 & LL & DHF
     & ZORA-HF & ZORA-DFT-D3 & LL & DHF\\ \hline
Bk--N$_1$&2.521 & 2.338 & 2.494 & 2.547 & 2.585 & 0.183 & -0.027 & 0.026 &0.064 \\
Bk--N$_2$&2.571 & 2.406 & 2.441 & 2.432 & 2.584 & 0.165 & -0.13 & -0.139 &-0.013 \\
Bk--N$_3$&2.544 & 2.354 & 2.492 & 2.486 & 2.587 & 0.178 & -0.04 & -0.046 &0.055 \\
RMSD     &      &       &       &       &       & 0.175 &  0.08 & 0.086  &0.049 \\
         &      &       &       &       &       &       &       &        &  \\
Bk--O$_1$&2.459 &  2.346 & 2.59 & 2.579 & 2.478 & 0.114 & 0.13 & 0.119 & 0.018 \\
Bk--O$_2$&2.436 &  2.32 & 2.495 & 2.592 & 2.509 & 0.126 & 0.049 & 0.146&0.063 \\
Bk--O$_3$&2.454 &  2.336 & 2.505 & 2.569 & 2.492 & 0.117 & 0.052 & 0.116&0.039 \\
Bk--O$_4$&2.468 &  2.323 & 2.583 & 2.569 & 2.529 & 0.145 & 0.115 & 0.101&0.061 \\
Bk--O$_5$&2.447 &  2.329 & 2.532 & 2.555 & 2.534 & 0.118 & 0.085 & 0.108&0.087 \\
Bk--O$_6$& 2.452&  2.313 & 2.528 & 2.626 & 2.549 & 0.139 & 0.076 & 0.174&0.097 \\ \hline
RMSD     &       &       &       &       &       & 0.127 & 0.09  & 0.13 &0.066 \\ \hline
\end{tabular}}
\label{t:2}
\end{table}
$^{(a)}$ Taken from Ref. \citep{galley2019synthesis}.

The last system studied was Cf--DOPO for which the oxidation state is +3. Its equilibrium structure is shown in Figure \ref{f:1}d. The bond distances at equilibrium, Cf--N and Cf--O, are collected in Table S6. There are no experimental nor previous theoretical results for comparison of Cf complexes. Our calculations at the ZORA--DFT--D3 level of theory shows that the average Cf--N bond distance is slightly higher than the average Pu--N bond distance by 0.037 \AA, and similarly the average Cf--O bond distance is slightly higher than the average Pu--O bond distance by 0.083 \AA. The average of bond distances Cf--N and Cf--O are reported in Table S7, where, as happens with Am, the introduction of dispersion correction to the ZORA calculations changes the bond length slightly.

%
\subsection{Bonding energies}
%
For these calculations we shall consider the Am and Bk containing systems as model compounds. The bonding energies (BE) 
are calculated, at DHF and LL levels of approach, as: BE =  Ac[DOPO]$_3$ - (Ac(III) + 3DOPO$^{-1}$). Instead, at ZORA level the BE is calculated as the difference of energies among the whole molecule and its fragments. The main results are given in Table \ref{table:BEAm} (see also Tables 10-17 of SI). 

We have performed calculations at four-component levels in such a way that we are able to analize the influence of relativistic effects (DHF - LL), SO effects at DHF level of approach (DHF - DHF-SF) and electron correlation (4c-B3LYP - DHF). We have also performed calculations at ZORA level of theory and there we can analyze the influence of SO effects and the electron correlation. Given that we include calculations that consider the actinide atom with different multiplicities we are able to learn about what kind of calculations give results that are closer to the four-component ones.

\medskip

The behavior of relativistic effects is opposite for Am--(DOPO)$_3$ and Bk--(DOPO)$_3$ complexes. For the first actinide relativistic effects on BE are positive (0.139 au) of which SO effects are the most important (0.105 au). On the contrary, for berkelium relativistic effects on BE are of opposite sign (-0.319 au) and in this case the SO effect is smaller (-0.083 au) when it is compared with its equivalent in Am.

On the other hand, at sr-ZORA level of approach we found that BEs are much dependent on the multiplicity of the actinide atom. One should compare results of DHF-SF calculations with that of sr-ZORA/HF. For both actinides the more approximated results of ZORA calculations are found for the quintuplet multiplicity. Besides, given that the so-ZORA calculations are performed with the closed-shell system its results must be comparable with that of DHF.

\medskip

When electron correlation is considered at each of both levels of theory we found different behaviors. At ZORA level the electron correlation for Am is equal to: -2.128 au + 1.335 au = -0.793 au. For Bk it is -2362 + 1.850 = -0.712. In the case of 4c calculations, for Am the electron correlation is: -2.097 au + 1.475 au = -0.622 au which is close to that of ZORA. But in the case of Bk the electron correlation is: -2.121 au + 1.784 au = -0.337 au which is almost halve of that of the ZORA calculation.

Furthermore one may assume that when electron correlation effects are included at 4c-B3LYP level of theory their results must be close to those of so-ZORA/B3LYP calculations. In the case of Am-DOPO compound this is what we found: -2.097 au vs -2.128 au (-5505.67 kJ/mol vs -5587.06 kJ/mol) but for Bk-DOPO compound a little large difference appears: -2.121 au vs -2.362 au (-5568.69 kJ/mol vs -6201.43 kJ/mol). We shall be aware that both methodologies, ZORA/B3LYP and 4c/B3LYP have different grounds. The first one use frozen core description of the electronic behavior but the other is an all electron methodology. 

We could shed some more light on the reason why those differences arises by observing the energy-spectra of the frontier orbitals shown in Figure \ref{f:2}) for the Am--(DOPO)$_3$ complex. We show the energy of MOs that belongs to HOMO-LUMO structures of sr-ZORA/B3LYP, 4c/B3LYP and 4c-DHF. There is a clear difference among the energy-gap between HOMO-LUMO orbitals and also the influence of f-type AOs in them.

Another finding is that by including electron correlation both molecular systems become more stable.

\medskip

Our previous analysis shows that higher levels of theory, such as DHF, might be needed to obtain a better estimation not only of the geometry but also a more accurate estimation of reaction energies. Additionally, the basis set superposition error is less than 6\% for the BE of both, Am--(DOPO)$_3$ and Bk--(DOPO)$_3$ complexes.
\begin{table}[htbp]
\caption{Bonding energies (BE) for Am and Bk, calculated at DHF, DHF-spin-free (DHF-SF), L\'{e}vy-Leblond (LL) and sr (so)-ZORA levels of approach, expressed in a.u. (between parentheses in kJ/mol).}
\adjustbox{max width=\textwidth}{
\begin{tabular}{|lccccc|} \hline
&&&&& \\
                        & Am(III).3DOPO$^{-1}$ & Am(III) & DOPO$^{-1}$ & BE$^{a}$ & $\Delta$E$^{b}$ \\ 
DHF                     & -32694.990  & -30488.849 & -734.889 & -1.475 (-3872.61)& -0.0536 (140.81) \\
DHF-SF                  & -32619.097  & -30412.630 & -734.929 & -1.680 (-4410.80)&                  \\
LL                      & -29897.818  & -27692.640 & -734.521 & -1.614 (-4237.56)& -0.0878 (-230.45) \\
4c-B3LYP                & -32722.070  & -30503.042 & -738.977 & -2.097 (-5505.67)&                  \\
&&&&& \\
so-ZORA/HF              & -32.023     & -0.056      & -10.211  & -1.335 (-3505.04)& \\
sr-ZORA/HF$^{c}$        & -31.835     &  1.229      & -10.211  & -2.432 (-6385.22)& \\
sr-ZORA/HF$^{d}$        & -31.474     &  0.569      & -10.211  & -1.411 (-3704.58)& \\
sr-ZORA/B3LYP$^{c}$     & -20.814     &  1.457      &  -6.609  & -2.445 (-6418.56)& \\
sr-ZORA/B3LYP$^{d}$     & -20.756     &  1.296      &  -6.609  & -2.226 (-5844.36)& \\
sr-ZORA/B3LYP$^{e}$     & -20.593     &  1.096      &  -6.609  & -1.863 (-4890.08)& \\
so-ZORA/B3LYP           & -20.997     &  1.001      &  -6.609  & -2.128 (-5587.06)& \\
                        &             &            &          &                  &  \\
                        & Bk(III).3DOPO$^{-1}$ & Bk(III)    & DOPO$^{-1}$ & BE$^{a}$ & $\Delta$E$^{b}$ \\
DHF                     & -34397.417 & -32190.967 & -734.889 & -1.784 (-4683.89) & -0.060 (158.38) \\
DHF-SF                  & -34308.076 & -32101.587 & -734.929 & -1.701 (-4465.98)&                  \\
LL                      & -31301.618 & -29096.589 & -734.521 & -1.465 (-3846.36) & -0.105 (-274.53) \\
4c-B3LYP                & -34425.524 & -32206.471 & -738.977 & -2.121 (-5568.69)&                  \\
&&&&& \\
so-ZORA/HF              & -32.172    & 0.310      & -10.211  & -1.850 (-4857.20)  &\\
sr-ZORA/HF$^{c}$        & -31.916    & 1.377      & -10.211  & -2.661 (-6986.46) &\\
sr-ZORA/HF$^{d}$        & -31.707    & 0.780      & -10.211  & -1.855 (-4870.73) &\\
sr-ZORA/B3LYP$^{c}$     & -20.846    & 1.562      & -6.609   & -2.581 (-6776.42) &\\ 
sr-ZORA/B3LYP$^{d}$     & -20.804    & 1.390      & -6.609   & -2.367 (-6215.83) &\\
sr-ZORA/B3LYP$^{e}$     & -20.624    & 1.176      & -6.609   & -1.974 (-5182.32) &\\
so-ZORA/B3LYP           & -21.004    & 1.185      & -6.609   & -2.362 (-6201.43) &\\
&&&&& \\
\hline
\end{tabular}
\label{table:BEAm}
}
$^{a}$ At DHF and LL levels of approach: BE =  Ac[DOPO]$_3$ - (Ac(III) + 3DOPO$^{-1}$); at ZORA level, BE is calculated as the difference of energies among the whole molecule and its fragments.\\
$^{b}$ $\Delta E$ = E(DHF with optimized ZORA geometry) - E(DHF with optimized DHF geometry) or E(LL with ZORA geometry) - E(LL with LL optimized geometry), respectively.\\
$^{c}$ Singlet multiplicity.\\
$^{d}$ Quintuplet multiplicity.\\
$^{e}$ Septuplet multiplicity.\\

\end{table}
\subsection{Influence of p-, d- and f- AOs on the HOMO-LUMO electronic structure}
%
After getting the optimized structures of the Pu--(DOPO)$_3$, Am--(DOPO)$_3$, Cf--(DOPO)$_3$ and Bk--(DOPO)$_3$ complexes, we have further carried out an investigation of the contribution of the valence AOs of the heavy elements (Pu, Am, Cf, and Bk) to the frontier MOs by analyzingthe pattern of the HOMO--LUMO electronic structure and the LFD diagram. We got reliable results for both, Am and Bk compounds at 4c level of approach (Table S8 and Table S9). The other compounds are not closed--shell and so they cannot be studied with the current version of the DIRAC code. On the other hand, the Lévy-Leblond calculation of the structure of frontier orbitals of Bk--(DOPO)$_3$ is not as reliable for Am--(DOPO)$_3$. In this last case the HOMO - LUMO orbitals of Bk--(DOPO)$_3$ have some inconsistencies in the LUMO structure, likely arising from linear dependence.


%
\begin{figure}[htbp]
 \centering{
    \includegraphics[scale=0.55]{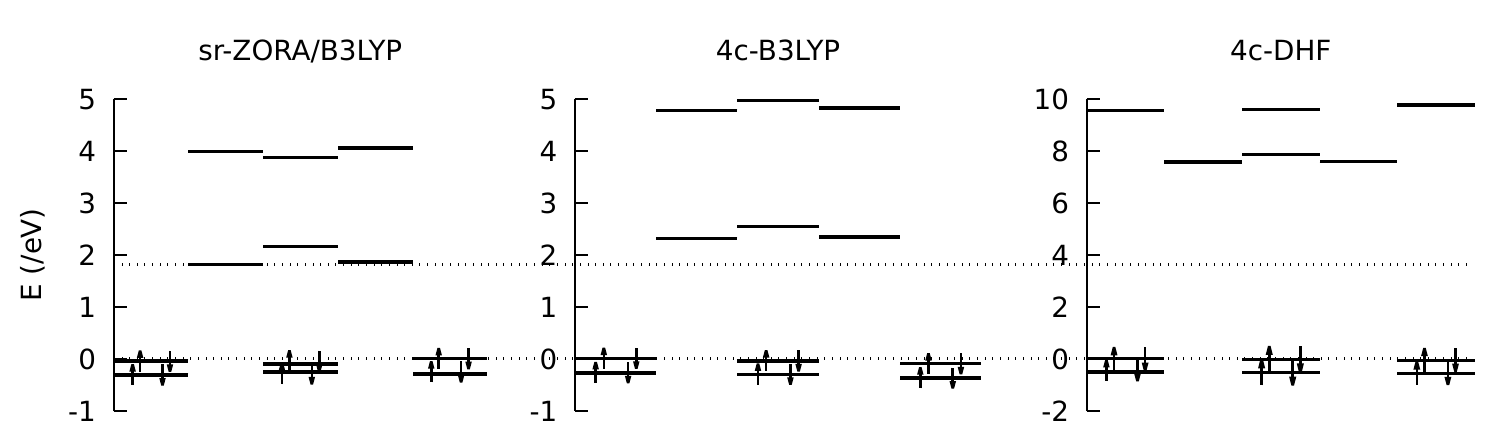}
 }
 \caption{Ligand Field Diagram (LFD) of the Am--DOPO system calculated at the ZORA, DHF and LL levels of theory. Green lines represents the OM with f-type primitives of the Am with significant Mulliken population.}
 \label{f:2}
\end{figure}

To further understand the bonding structure of the electronic framework in the Am(DOPO)$_3$ complex, LFD computations were first performed at the ZORA level (Figure \ref{f:2}). This calculation shows that the 5f--AOs of Am and DOPO ligands are taking part to create MOs with bonding character. The HOMO and LUMO orbitals in which those AOs are more involved are HOMO - 9, HOMO - 11, LUMO + 3 and LUMO + 6 (Figures S6 and S7). The study of 5f-- and 5p--AOs of Am contributions to frontier MOs was also performed at both, the DHF and LL levels of theory to corroborate the previous findings at the ZORA level. There are significant contributions of Am 5f-- and 6p--AOs to HOMO - 6 (fxyy = 0.271, fxxy = 0.271, fxxx = 0.091) and LUMO + 3 (py = 0.633, px = 0.241, pz = 0.043) molecular orbitals (Table S9). Thus, we found a similar scheme to that of ZORA--DFT--D3 though with little differences (see Figure \ref{f:2}). It is important to point out that those relativistic effects are largely involved. At the LL level of theory, the contribution of 5f--AO of Am to HOMO--6 is lost, and there is no significant contribution of them to any HOMO - n (n = 1 - 12). On the other hand, the pattern of the 5f--AOs contributions to LUMO + n orbitals (n = 9) are just slightly modified, e.g., LUMO + 3 (py = 0.587, px = 0.234, pz = 0.043). Thus, without the relativistic approximations (the LL reference method), there are no contributions to the HOMO - n orbitals, which are present when ZORA or DHF are used. This is a remarkable finding because it implies that relativistic effects play a central role in the involvement of f--orbitals in the frontier orbitals. Intuitively, this is an expected result due to the well--known influence of direct/indirect relativistic effects on the size of AOs, which means the contraction/expansion of the p--AOs/f--AOs. This is now numerically observed on these actinide compounds. Furthermore, the pattern of energies of the set of HOMO - n (n= 1 - 8) orbitals at both, four--component and NR level of theory are quite similar.

A similar interaction symmetry can be seen in LUMO + 4 and LUMO + 5.


%
\begin{figure}[h!]
 \centering{
    \includegraphics[scale=0.55]{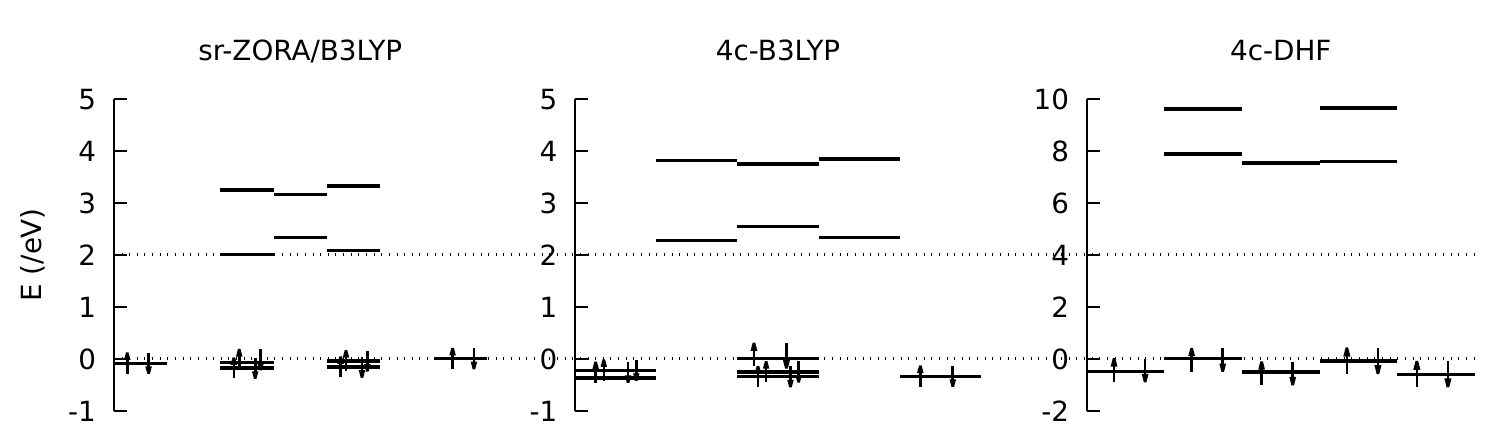}
 }
 \caption{Ligand Field Diagram (LFD) of the Bk-DOPO system calculated at the ZORA, DHF and LL levels of theory. The lines in green represents the OM with f-type primitives of the Bk with significant Mulliken magnitudes.}
 \label{f:3}
\end{figure}

 The analysis of the contributions from Bk 5f--AOs to the frontier MOs was also performed at the LL and DHF level of theory for comparison with ZORA calculations. There are significant contributions of Bk 5f-- and 6p--AOs to HOMO - 8 (fxyy = 0.171, fxxy = 0.114, fxxx = 0.063) and LUMO + 3 (py = 0.578, px = 0.268, fxyy = 0.054) molecular orbitals at the DHF level (Table S10). The non--relativistic calculations were performed with LL Hamiltonian, but in this case some inconsistencies appears because the LUMO and LUMO + 1 orbital energies are negative, and few linear dependencies arises in the wave function calculation. In summary, as was also observed for the complexed Am compound the symmetry, magnitude and types of Bk--AOs are affected by relativistic effects, which in turn will affect how we interpret bonding and reaction on these compounds. This calls for caution when interpreting standard DFT calculations without scalar or 4--component relativistic effects for bonding or reactions.

%
%
First, we focus on the Pu AOs involved in the bonding. The LFD of the Pu--(DOPO)$_3$ complex shows that the electronic 5f--orbitals of Plutonium are interacting with the 2p--orbitals of both, the N and O atoms (Figure S3). To provide further insight about the interaction between Pu and N or O atoms, we have computed the HOMO - LUMO and HOMO - n and LUMO + n MOs (Figure S4 and S5). These diagrams shows that the Pu AOs interact with the DOPO  ligand AOs to create molecular orbitals with bonding character. To give an example, three of the highest bonding molecular orbitals, HOMO, HOMO - 6 and HOMO - 9 show that 5f--orbital electrons of Plutonium interact with the 2p--AOs of the N, O and even C atoms. More specifically in the HOMO, the 5f--AO is delocalized with a 2p AO from N atom in the organic ligand (Figure S4). Similarly, we can observe in LUMO, LUMO + 1, LUMO + 2 and LUMO + 6 that the Pu 5f--AOs are delocalized with the 2p--AO from N and O atoms, but not C, in the ligand (Figure S5). This is evidence that at the ZORA level of theory the 5f--AOs of the Pu plays an important role in creating a bond with the organic ligand, i.e., 5f-AOs are crucial to describe the bonding of the Pu(DOPO)$_3$ complex.

\medskip

The bonding interactions between Cf and DOPO (Figure S8) was studied by analyzing the LFD of Cf--(DOPO)$_3$. The LFD shows that near the frontier orbitals, the 5f--AOs of Cf interact with the N, O and even with the C atoms in the HOMO orbitals. However only C and N participate in the LUMO near the frontier. The HOMO and LUMO orbitals were calculated at the ZORA level of theory (Figures S9-10). In HOMO--7, HOMO--8 and HOMO--10, the overlap between 5f--AOs and AOs from the ligand can be observed, which have some bonding character. A similar study was performed on Bk containing compounds. The LFD shows that 5f--AOs of Bk are interacting with the carbon atom in the DOPO ligand of the complex (Figure \ref{f:3} and S11), which is also observed qualitatively in the HOMO - LUMO representations at the ZORA level (Figures S12-13). The overlap of the 5f--AOs can be seen in HOMO - 7, HOMO - 8 and HOMO - 9. 

%
\section{Conclusions} \label{sec:conclusions}
%
We computationally investigated the molecular structure at equilibrium of the f-block element (Pu, Am, Bk, and Cf) compounds of the redox-active DOPO ligand by using i) four--component with Dirac--Hartree--Fock (DHF), Lévy-Leblond (LL) and spin-free (SF) Hamiltonians, together with ii) the scalar relativistic (sr) and spin-orbit relativistic (so) zeroth--order regular approximation (ZORA) with hybrid density functional theory (DFT).

\medskip

We found that the RMSD values for bond distances are the smallest when the DHF method is applied, followed by results of calculations with ZORA and LL levels of theory. Taking both $Ac$--DOPO ($Ac$ = Am, Bk) compounds as model compounds, we found that relativistic effects shorten the bond distances on average by 0.04\AA~for Am--O and 0.07\AA~for Bk--O. Furthermore, all six Am--O bond distances and the three Am--N bond distances become closer to each other when four--component relativistic effects are included, and they are also closer to experimental distances as compared with sr-ZORA results. Total relativistic effects shortened the $Ac$--O bond distances but expand the $Ac$--N bond distances. When considering the scalar relativistic effects, we found that they shorten $Ac$--N bonds lengths but the spin--dependent relativistic effects expand them. The opposite behavior is found for the $Ac$--O bond lengths. Furthermore, electron correlation stretches bond distances of $Ac$--N and $Ac$--O though the effect on stretching the distance d($Ac$--O) is larger than that effect on d($Ac$--N).

\medskip

On the other hand, we found that the BE is more negative when it is calculated at 4c-DHF level of theory than for so-ZORA/HF level of theory for Am-DOPO$_3$ complex (-3872.61 kJ/mol vs -3505.04 kJ/mol, respectively). For the Bk-DOPO$_3$ complex such a tendence is opposite (-4683.89 kJ/mol vs -4857.20 kJ/mol, respectively). In both cases the inclusion of electron correlation at any level of theory makes those systems more stables. This illustrates that working with more accurate levels of theory, such as the one that consider four-component Hamiltonians and all-electron calculations, might be needed to obtain a better estimation of the geometrical structure and even a more accurate estimation of reaction energies for $Ac$-DOPO$_3$ complexes.

We have also analized the contributions of 5f-- and 6p--AOs of central actinides ($Ac$ = Am, Bk) to frontier MOs at both levels of theory, ZORA and DHF. We found that for $Ac$ = Am, there are significant contributions of Am 5f-- and 6p--AOs to the HOMO - 6 (fxyy = 0.271, fxxy = 0.271, fxxx = 0.091) and LUMO + 3 (py = 0.633, px = 0.241, pz = 0.043). However, with the non-relativistic LL Hamiltonian the Am 5f--AO contribution to HOMO - 6 is lost and also any other HOMO - n (n = 1-12). This means that, without including relativistic effects at ZORA or DHF level, no contributions of the Am 5f--AOs to the HOMO - n orbitals are found. This shows that relativistic effects play a key role in the population of f--orbitals in the frontier orbitals. A similar behavior occurs for the Bk--containing compound, where also the symmetry and the magnitude of Bk--AOs are affected by relativistic effects. In this case the HOMO - 8 (fxyy = 0.171, fxxy = 0.114, fxxx = 0.063) and LUMO + 3 (py = 0.578, px = 0.268, pz = 0.054) have Bk-AO contributions.

\medskip

We can state now that calculations at both, ZORA and DHF (with a given local dense basis set) levels of approach give good estimation of geometries, which can be useful when the experimental synthesis is challenging or not cost-effective, and also that the reaction of these complexes can only be adequately reproduced if they are described within a relativistic framework, given that the change depends on the frontier MOs. This is a call for caution for using standard DFT calculations when the complex studied involves actinide elements. One must include either, scalar or four--component relativistic effects on the calculations of bondings and reactions. In addition, we want to emphasize that one must include an adequate local dense basis set scheme for making that results of calculations at DHF level be reliable.

As a continuation of our line of research and based on these results, we started the study of some response properties like the NMR spectroscopic parameters, which shall proportion new insights about the physics that is behind these Actinide--containing systems.
\section{Supporting Information} \label{sec:Supporting Information}
The equilibrium geometries with the x, y, z coordinators using the scalar (ZORA) and the 4--component DHF and LL Hamiltonian are given in the Supporting Information.
\begin{acknowledgement}
G.A.A. acknowledge support from the Argentinian National Founding for Science and Technique, FONCYT (Grant PICT2016-2936). S.P. thanks the Science and Engineering Research Board, Department of Science and Technology (SERB-DST), Govt. of India for providing the Ramanujan Faculty Fellowship under the scheme no. SB/S2/RJN-067/2017 and the Early Career Research Award (ECRA) under the project number ECR/2018/000255. The computing for this project was performed partly on the High-Performance Computer cluster at the Research Computing Center at Florida State University (FSU). The large portion of this work was supported in part through computational resources and services provided by the Institute for Cyber--Enabled Research at Michigan State University (MSU).
\end{acknowledgement}
\section{Conflicts of interest} \label{sec:Conflicts of interest}
There are no conflicts to declare.
%
\bibliography{references}

\providecommand{\latin}[1]{#1}
\makeatletter
\providecommand{\doi}
  {\begingroup\let\do\@makeother\dospecials
  \catcode`\{=1 \catcode`\}=2 \doi@aux}
\providecommand{\doi@aux}[1]{\endgroup\texttt{#1}}
\makeatother
\providecommand*\mcitethebibliography{\thebibliography}
\csname @ifundefined\endcsname{endmcitethebibliography}
  {\let\endmcitethebibliography\endthebibliography}{}
\begin{mcitethebibliography}{39}
\providecommand*\natexlab[1]{#1}
\providecommand*\mciteSetBstSublistMode[1]{}
\providecommand*\mciteSetBstMaxWidthForm[2]{}
\providecommand*\mciteBstWouldAddEndPuncttrue
  {\def\EndOfBibitem{\unskip.}}
\providecommand*\mciteBstWouldAddEndPunctfalse
  {\let\EndOfBibitem\relax}
\providecommand*\mciteSetBstMidEndSepPunct[3]{}
\providecommand*\mciteSetBstSublistLabelBeginEnd[3]{}
\providecommand*\EndOfBibitem{}
\mciteSetBstSublistMode{f}
\mciteSetBstMaxWidthForm{subitem}{(\alph{mcitesubitemcount})}
\mciteSetBstSublistLabelBeginEnd
  {\mcitemaxwidthsubitemform\space}
  {\relax}
  {\relax}

\bibitem[Y.~Ohtsuka(2019)]{Ohtsuka2019}
Y.~Ohtsuka,~Y. T. Y. I. M. H. K. H. S.~H.,~M.~Aoyama \emph{Sci. Rep.}
  \textbf{2019}, \emph{9}, 1--8\relax
\mciteBstWouldAddEndPuncttrue
\mciteSetBstMidEndSepPunct{\mcitedefaultmidpunct}
{\mcitedefaultendpunct}{\mcitedefaultseppunct}\relax
\EndOfBibitem
\bibitem[Greenwood and Earnshaw(1997)Greenwood, and
  Earnshaw]{greenwood1997chemistry}
Greenwood,~N.; Earnshaw,~A. \emph{Chemistry of the Elements 2nd Edition};
  Butterworth-Heinemann, 1997\relax
\mciteBstWouldAddEndPuncttrue
\mciteSetBstMidEndSepPunct{\mcitedefaultmidpunct}
{\mcitedefaultendpunct}{\mcitedefaultseppunct}\relax
\EndOfBibitem
\bibitem[Galley \latin{et~al.}(2019)Galley, Pattenaude, Gaggioli, Qiao,
  Sperling, Zeller, Pakhira, Mendoza-Cortes, Schelter, Albrecht-Schmitt,
  \latin{et~al.} others]{galley2019synthesis}
Galley,~S.~S.; Pattenaude,~S.~A.; Gaggioli,~C.~A.; Qiao,~Y.; Sperling,~J.~M.;
  Zeller,~M.; Pakhira,~S.; Mendoza-Cortes,~J.~L.; Schelter,~E.~J.;
  Albrecht-Schmitt,~T.~E., \latin{et~al.}  Synthesis and characterization of
  tris-chelate complexes for understanding f-orbital bonding in later
  actinides. \emph{Journal of the American Chemical Society} \textbf{2019},
  \emph{141}, 2356--2366\relax
\mciteBstWouldAddEndPuncttrue
\mciteSetBstMidEndSepPunct{\mcitedefaultmidpunct}
{\mcitedefaultendpunct}{\mcitedefaultseppunct}\relax
\EndOfBibitem
\bibitem[Brenner \latin{et~al.}(2020)Brenner, Sperling, Poe, Celis-Barros,
  Brittain, Villa, Albrecht-Schmitt, and Polinski]{brenner2020trivalent}
Brenner,~N.; Sperling,~J.~M.; Poe,~T.~N.; Celis-Barros,~C.; Brittain,~K.;
  Villa,~E.~M.; Albrecht-Schmitt,~T.~E.; Polinski,~M.~J. Trivalent f-element
  squarates, squarate-oxalates, and cationic materials, and the determination
  of the nine-coordinate ionic radius of Cf (III). \emph{Inorganic Chemistry}
  \textbf{2020}, \emph{59}, 9384--9395\relax
\mciteBstWouldAddEndPuncttrue
\mciteSetBstMidEndSepPunct{\mcitedefaultmidpunct}
{\mcitedefaultendpunct}{\mcitedefaultseppunct}\relax
\EndOfBibitem
\bibitem[Leszczyk \latin{et~al.}(2019)Leszczyk, Tecmer, and
  Boguslawski]{leszczyk2019new}
Leszczyk,~A.; Tecmer,~P.; Boguslawski,~K. \emph{Transition Metals in
  Coordination Environments}; Springer, 2019; pp 121--160\relax
\mciteBstWouldAddEndPuncttrue
\mciteSetBstMidEndSepPunct{\mcitedefaultmidpunct}
{\mcitedefaultendpunct}{\mcitedefaultseppunct}\relax
\EndOfBibitem
\bibitem[Pyykk{\"o}(2012)]{pyykko2012relativistic}
Pyykk{\"o},~P. Relativistic effects in chemistry: more common than you thought.
  \emph{Annual review of physical chemistry} \textbf{2012}, \emph{63},
  45--64\relax
\mciteBstWouldAddEndPuncttrue
\mciteSetBstMidEndSepPunct{\mcitedefaultmidpunct}
{\mcitedefaultendpunct}{\mcitedefaultseppunct}\relax
\EndOfBibitem
\bibitem[Chang \latin{et~al.}(1986)Chang, Pelissier, and
  Durand]{chang1986regular}
Chang,~C.; Pelissier,~M.; Durand,~P. Regular two-component Pauli-like effective
  Hamiltonians in Dirac theory. \emph{Physica Scripta} \textbf{1986},
  \emph{34}, 394\relax
\mciteBstWouldAddEndPuncttrue
\mciteSetBstMidEndSepPunct{\mcitedefaultmidpunct}
{\mcitedefaultendpunct}{\mcitedefaultseppunct}\relax
\EndOfBibitem
\bibitem[van Lenthe \latin{et~al.}(1994)van Lenthe, Baerends, and
  Snijders]{van1994relativistic}
van Lenthe,~E.; Baerends,~E.-J.; Snijders,~J.~G. Relativistic total energy
  using regular approximations. \emph{The Journal of chemical physics}
  \textbf{1994}, \emph{101}, 9783--9792\relax
\mciteBstWouldAddEndPuncttrue
\mciteSetBstMidEndSepPunct{\mcitedefaultmidpunct}
{\mcitedefaultendpunct}{\mcitedefaultseppunct}\relax
\EndOfBibitem
\bibitem[Saue(2011)]{saue2011relativistic}
Saue,~T. Relativistic Hamiltonians for chemistry: A primer. \emph{ChemPhysChem}
  \textbf{2011}, \emph{12}, 3077--3094\relax
\mciteBstWouldAddEndPuncttrue
\mciteSetBstMidEndSepPunct{\mcitedefaultmidpunct}
{\mcitedefaultendpunct}{\mcitedefaultseppunct}\relax
\EndOfBibitem
\bibitem[A.~Goel(2019)]{goel2019}
A.~Goel,~R. P. A. A.~K.,~J. S.~McCloy \emph{J. Non--Cryst. Sol.: X}
  \textbf{2019}, \emph{4}, 100033\relax
\mciteBstWouldAddEndPuncttrue
\mciteSetBstMidEndSepPunct{\mcitedefaultmidpunct}
{\mcitedefaultendpunct}{\mcitedefaultseppunct}\relax
\EndOfBibitem
\bibitem[Vienna(2010)]{vienna2010nuclear}
Vienna,~J.~D. Nuclear waste vitrification in the United States: recent
  developments and future options. \emph{International Journal of Applied Glass
  Science} \textbf{2010}, \emph{1}, 309--321\relax
\mciteBstWouldAddEndPuncttrue
\mciteSetBstMidEndSepPunct{\mcitedefaultmidpunct}
{\mcitedefaultendpunct}{\mcitedefaultseppunct}\relax
\EndOfBibitem
\bibitem[Fisk \latin{et~al.}(1988)Fisk, Hess, Pethick, Pines, Smith, Thompson,
  and Willis]{fisk1988heavy}
Fisk,~Z.; Hess,~D.; Pethick,~C.; Pines,~D.; Smith,~J.; Thompson,~J.; Willis,~J.
  Heavy-electron metals: New highly correlated states of matter. \emph{Science}
  \textbf{1988}, \emph{239}, 33--42\relax
\mciteBstWouldAddEndPuncttrue
\mciteSetBstMidEndSepPunct{\mcitedefaultmidpunct}
{\mcitedefaultendpunct}{\mcitedefaultseppunct}\relax
\EndOfBibitem
\bibitem[van Schilfgaarde \latin{et~al.}(1999)van Schilfgaarde, Abrikosov, and
  Johansson]{van1999origin}
van Schilfgaarde,~M.; Abrikosov,~I.; Johansson,~B. Origin of the Invar effect
  in iron--nickel alloys. \emph{Nature} \textbf{1999}, \emph{400}, 46--49\relax
\mciteBstWouldAddEndPuncttrue
\mciteSetBstMidEndSepPunct{\mcitedefaultmidpunct}
{\mcitedefaultendpunct}{\mcitedefaultseppunct}\relax
\EndOfBibitem
\bibitem[Boring and Smith(2000)Boring, and Smith]{boring2000plutonium}
Boring,~A.~M.; Smith,~J.~L. Plutonium condensed-matter Physics. \emph{Los
  Alamos Science} \textbf{2000}, \emph{26}, 90\relax
\mciteBstWouldAddEndPuncttrue
\mciteSetBstMidEndSepPunct{\mcitedefaultmidpunct}
{\mcitedefaultendpunct}{\mcitedefaultseppunct}\relax
\EndOfBibitem
\bibitem[Cao \latin{et~al.}(2021)Cao, Vernon, Schwarz, and
  Li]{cao2021understanding}
Cao,~C.; Vernon,~R.~E.; Schwarz,~W.; Li,~J. Understanding periodic and
  non-periodic chemistry in periodic tables. \emph{Frontiers in Chemistry}
  \textbf{2021}, \emph{8}, 813\relax
\mciteBstWouldAddEndPuncttrue
\mciteSetBstMidEndSepPunct{\mcitedefaultmidpunct}
{\mcitedefaultendpunct}{\mcitedefaultseppunct}\relax
\EndOfBibitem
\bibitem[Cooper(2000)]{cooper2000challenges}
Cooper,~N.~G. Challenges in plutonium science. \emph{Los Alamos Sci}
  \textbf{2000}, \emph{26}, 1--493\relax
\mciteBstWouldAddEndPuncttrue
\mciteSetBstMidEndSepPunct{\mcitedefaultmidpunct}
{\mcitedefaultendpunct}{\mcitedefaultseppunct}\relax
\EndOfBibitem
\bibitem[Galley \latin{et~al.}(2020)Galley, Gaggioli, Zeller, Celis-Barros,
  Albrecht-Schmitt, Gagliardi, and Bart]{galley2020evidence}
Galley,~S.~S.; Gaggioli,~C.~A.; Zeller,~M.; Celis-Barros,~C.;
  Albrecht-Schmitt,~T.~E.; Gagliardi,~L.; Bart,~S.~C. Evidence of Alpha
  Radiolysis in the Formation of a Californium Nitrate Complex.
  \emph{Chemistry--A European Journal} \textbf{2020}, \emph{26},
  8885--8888\relax
\mciteBstWouldAddEndPuncttrue
\mciteSetBstMidEndSepPunct{\mcitedefaultmidpunct}
{\mcitedefaultendpunct}{\mcitedefaultseppunct}\relax
\EndOfBibitem
\bibitem[Tobisu \latin{et~al.}(2012)Tobisu, Kinuta, Kita, Remond, and
  Chatani]{tobisu2012rhodium}
Tobisu,~M.; Kinuta,~H.; Kita,~Y.; Remond,~E.; Chatani,~N. Rhodium (I)-catalyzed
  borylation of nitriles through the cleavage of carbon--cyano bonds.
  \emph{Journal of the American Chemical Society} \textbf{2012}, \emph{134},
  115--118\relax
\mciteBstWouldAddEndPuncttrue
\mciteSetBstMidEndSepPunct{\mcitedefaultmidpunct}
{\mcitedefaultendpunct}{\mcitedefaultseppunct}\relax
\EndOfBibitem
\bibitem[Kundu \latin{et~al.}(2009)Kundu, Choliy, Zhuo, Ahuja, Emge, Warmuth,
  Brookhart, Krogh-Jespersen, and Goldman]{kundu2009rational}
Kundu,~S.; Choliy,~Y.; Zhuo,~G.; Ahuja,~R.; Emge,~T.~J.; Warmuth,~R.;
  Brookhart,~M.; Krogh-Jespersen,~K.; Goldman,~A.~S. Rational design and
  synthesis of highly active pincer-iridium catalysts for alkane
  dehydrogenation. \emph{Organometallics} \textbf{2009}, \emph{28},
  5432--5444\relax
\mciteBstWouldAddEndPuncttrue
\mciteSetBstMidEndSepPunct{\mcitedefaultmidpunct}
{\mcitedefaultendpunct}{\mcitedefaultseppunct}\relax
\EndOfBibitem
\bibitem[Silver \latin{et~al.}(2016)Silver, Cary, Johnson, Baumbach, Arico,
  Luckey, Urban, Wang, Polinski, Chemey, \latin{et~al.}
  others]{silver2016characterization}
Silver,~M.~A.; Cary,~S.~K.; Johnson,~J.~A.; Baumbach,~R.~E.; Arico,~A.~A.;
  Luckey,~M.; Urban,~M.; Wang,~J.~C.; Polinski,~M.~J.; Chemey,~A.,
  \latin{et~al.}  Characterization of berkelium (III) dipicolinate and borate
  compounds in solution and the solid state. \emph{Science} \textbf{2016},
  \emph{353}, aaf3762\relax
\mciteBstWouldAddEndPuncttrue
\mciteSetBstMidEndSepPunct{\mcitedefaultmidpunct}
{\mcitedefaultendpunct}{\mcitedefaultseppunct}\relax
\EndOfBibitem
\bibitem[Albrecht-Schmitt \latin{et~al.}(2020)Albrecht-Schmitt, Hobart,
  P{\'a}ez-Hern{\'a}ndez, and Celis-Barros]{albrecht2020theoretical}
Albrecht-Schmitt,~T.~E.; Hobart,~D.~E.; P{\'a}ez-Hern{\'a}ndez,~D.;
  Celis-Barros,~C. Theoretical examination of covalency in berkelium (IV)
  carbonate complexes. \emph{International Journal of Quantum Chemistry}
  \textbf{2020}, \emph{120}, e26254\relax
\mciteBstWouldAddEndPuncttrue
\mciteSetBstMidEndSepPunct{\mcitedefaultmidpunct}
{\mcitedefaultendpunct}{\mcitedefaultseppunct}\relax
\EndOfBibitem
\bibitem[Goodwin \latin{et~al.}(2019)Goodwin, Su, Albrecht-Schmitt, Blake,
  Batista, Daly, Dehnen, Evans, Gaunt, Kozimor, \latin{et~al.}
  others]{goodwin2019back}
Goodwin,~C.~A.; Su,~J.; Albrecht-Schmitt,~T.~E.; Blake,~A.~V.; Batista,~E.~R.;
  Daly,~S.~R.; Dehnen,~S.; Evans,~W.~J.; Gaunt,~A.~J.; Kozimor,~S.~A.,
  \latin{et~al.}  Back Cover:[Am (C5Me4H) 3]: An Organometallic Americium
  Complex (Angew. Chem. Int. Ed. 34/2019). \emph{Angewandte Chemie
  International Edition} \textbf{2019}, \emph{58}, 11924--11924\relax
\mciteBstWouldAddEndPuncttrue
\mciteSetBstMidEndSepPunct{\mcitedefaultmidpunct}
{\mcitedefaultendpunct}{\mcitedefaultseppunct}\relax
\EndOfBibitem
\bibitem[Becke(1993)]{becke1993becke}
Becke,~A.~D. Becke’s three parameter hybrid method using the LYP correlation
  functional. \emph{J. Chem. Phys} \textbf{1993}, \emph{98}, 5648--5652\relax
\mciteBstWouldAddEndPuncttrue
\mciteSetBstMidEndSepPunct{\mcitedefaultmidpunct}
{\mcitedefaultendpunct}{\mcitedefaultseppunct}\relax
\EndOfBibitem
\bibitem[Lee \latin{et~al.}(1988)Lee, Yang, and Parr]{lee1988development}
Lee,~C.; Yang,~W.; Parr,~R.~G. Development of the Colle-Salvetti
  correlation-energy formula into a functional of the electron density.
  \emph{Physical review B} \textbf{1988}, \emph{37}, 785\relax
\mciteBstWouldAddEndPuncttrue
\mciteSetBstMidEndSepPunct{\mcitedefaultmidpunct}
{\mcitedefaultendpunct}{\mcitedefaultseppunct}\relax
\EndOfBibitem
\bibitem[Grimme(2006)]{grimme2006semiempirical}
Grimme,~S. Semiempirical GGA-type density functional constructed with a
  long-range dispersion correction. \emph{Journal of computational chemistry}
  \textbf{2006}, \emph{27}, 1787--1799\relax
\mciteBstWouldAddEndPuncttrue
\mciteSetBstMidEndSepPunct{\mcitedefaultmidpunct}
{\mcitedefaultendpunct}{\mcitedefaultseppunct}\relax
\EndOfBibitem
\bibitem[Lenthe \latin{et~al.}(1993)Lenthe, Baerends, and
  Snijders]{lenthe1993relativistic}
Lenthe,~E.~v.; Baerends,~E.-J.; Snijders,~J.~G. Relativistic regular
  two-component Hamiltonians. \emph{The Journal of chemical physics}
  \textbf{1993}, \emph{99}, 4597--4610\relax
\mciteBstWouldAddEndPuncttrue
\mciteSetBstMidEndSepPunct{\mcitedefaultmidpunct}
{\mcitedefaultendpunct}{\mcitedefaultseppunct}\relax
\EndOfBibitem
\bibitem[Van~Lenthe \latin{et~al.}(1996)Van~Lenthe, Van~Leeuwen, Baerends, and
  Snijders]{van1996relativistic}
Van~Lenthe,~E.; Van~Leeuwen,~R.; Baerends,~E.; Snijders,~J. Relativistic
  regular two-component Hamiltonians. \emph{International Journal of Quantum
  Chemistry} \textbf{1996}, \emph{57}, 281--293\relax
\mciteBstWouldAddEndPuncttrue
\mciteSetBstMidEndSepPunct{\mcitedefaultmidpunct}
{\mcitedefaultendpunct}{\mcitedefaultseppunct}\relax
\EndOfBibitem
\bibitem[Van~Lenthe \latin{et~al.}(1996)Van~Lenthe, Snijders, and
  Baerends]{van1996zero}
Van~Lenthe,~E.~v.; Snijders,~J.; Baerends,~E. The zero-order regular
  approximation for relativistic effects: The effect of spin--orbit coupling in
  closed shell molecules. \emph{The Journal of chemical physics} \textbf{1996},
  \emph{105}, 6505--6516\relax
\mciteBstWouldAddEndPuncttrue
\mciteSetBstMidEndSepPunct{\mcitedefaultmidpunct}
{\mcitedefaultendpunct}{\mcitedefaultseppunct}\relax
\EndOfBibitem
\bibitem[Cremer \latin{et~al.}(2014)Cremer, Zou, and Filatov]{cremer2014dirac}
Cremer,~D.; Zou,~W.; Filatov,~M. Dirac-exact relativistic methods: the
  normalized elimination of the small component method. \emph{Wiley
  Interdisciplinary Reviews: Computational Molecular Science} \textbf{2014},
  \emph{4}, 436--467\relax
\mciteBstWouldAddEndPuncttrue
\mciteSetBstMidEndSepPunct{\mcitedefaultmidpunct}
{\mcitedefaultendpunct}{\mcitedefaultseppunct}\relax
\EndOfBibitem
\bibitem[Van~Lenthe and Baerends(2003)Van~Lenthe, and
  Baerends]{van2003optimized}
Van~Lenthe,~E.; Baerends,~E.~J. Optimized Slater-type basis sets for the
  elements 1--118. \emph{Journal of computational chemistry} \textbf{2003},
  \emph{24}, 1142--1156\relax
\mciteBstWouldAddEndPuncttrue
\mciteSetBstMidEndSepPunct{\mcitedefaultmidpunct}
{\mcitedefaultendpunct}{\mcitedefaultseppunct}\relax
\EndOfBibitem
\bibitem[Guell \latin{et~al.}(2008)Guell, Luis, Sola, and
  Swart]{guell2008importance}
Guell,~M.; Luis,~J.~M.; Sola,~M.; Swart,~M. Importance of the basis set for the
  spin-state energetics of iron complexes. \emph{The Journal of Physical
  Chemistry A} \textbf{2008}, \emph{112}, 6384--6391\relax
\mciteBstWouldAddEndPuncttrue
\mciteSetBstMidEndSepPunct{\mcitedefaultmidpunct}
{\mcitedefaultendpunct}{\mcitedefaultseppunct}\relax
\EndOfBibitem
\bibitem[Dyall(2007)]{dyall2007relativistic}
Dyall,~K.~G. Relativistic double-zeta, triple-zeta, and quadruple-zeta basis
  sets for the actinides Ac--Lr. \emph{Theoretical Chemistry Accounts}
  \textbf{2007}, \emph{117}, 491--500\relax
\mciteBstWouldAddEndPuncttrue
\mciteSetBstMidEndSepPunct{\mcitedefaultmidpunct}
{\mcitedefaultendpunct}{\mcitedefaultseppunct}\relax
\EndOfBibitem
\bibitem[Dyall(2012)]{dyall2012core}
Dyall,~K.~G. Core correlating basis functions for elements 31--118.
  \emph{Theoretical Chemistry Accounts} \textbf{2012}, \emph{131}, 1--11\relax
\mciteBstWouldAddEndPuncttrue
\mciteSetBstMidEndSepPunct{\mcitedefaultmidpunct}
{\mcitedefaultendpunct}{\mcitedefaultseppunct}\relax
\EndOfBibitem
\bibitem[Binkley \latin{et~al.}(1980)Binkley, Pople, and
  Hehre]{binkley1980self}
Binkley,~J.~S.; Pople,~J.~A.; Hehre,~W.~J. Self-consistent molecular orbital
  methods. 21. Small split-valence basis sets for first-row elements.
  \emph{Journal of the American Chemical Society} \textbf{1980}, \emph{102},
  939--947\relax
\mciteBstWouldAddEndPuncttrue
\mciteSetBstMidEndSepPunct{\mcitedefaultmidpunct}
{\mcitedefaultendpunct}{\mcitedefaultseppunct}\relax
\EndOfBibitem
\bibitem[Gordon \latin{et~al.}(1982)Gordon, Binkley, Pople, Pietro, and
  Hehre]{gordon1982self}
Gordon,~M.~S.; Binkley,~J.~S.; Pople,~J.~A.; Pietro,~W.~J.; Hehre,~W.~J.
  Self-consistent molecular-orbital methods. 22. Small split-valence basis sets
  for second-row elements. \emph{Journal of the American Chemical Society}
  \textbf{1982}, \emph{104}, 2797--2803\relax
\mciteBstWouldAddEndPuncttrue
\mciteSetBstMidEndSepPunct{\mcitedefaultmidpunct}
{\mcitedefaultendpunct}{\mcitedefaultseppunct}\relax
\EndOfBibitem
\bibitem[Gomes \latin{et~al.}()Gomes, Saue, Visscher, Jensen, Bast, Aucar,
  Bakken, Dyall, Dubillard, Ekstr{\"o}m, \latin{et~al.} others]{gomesdirac19}
Gomes,~A.; Saue,~T.; Visscher,~L.; Jensen,~H.; Bast,~R.; Aucar,~I.; Bakken,~V.;
  Dyall,~K.; Dubillard,~S.; Ekstr{\"o}m,~U., \latin{et~al.}  DIRAC19.
  \emph{Zenodo} \relax
\mciteBstWouldAddEndPunctfalse
\mciteSetBstMidEndSepPunct{\mcitedefaultmidpunct}
{}{\mcitedefaultseppunct}\relax
\EndOfBibitem
\bibitem[Te~Velde \latin{et~al.}(2001)Te~Velde, Bickelhaupt, Baerends,
  Fonseca~Guerra, van Gisbergen, Snijders, and Ziegler]{te2001chemistry}
Te~Velde,~G.~t.; Bickelhaupt,~F.~M.; Baerends,~E.~J.; Fonseca~Guerra,~C.; van
  Gisbergen,~S.~J.; Snijders,~J.~G.; Ziegler,~T. Chemistry with ADF.
  \emph{Journal of Computational Chemistry} \textbf{2001}, \emph{22},
  931--967\relax
\mciteBstWouldAddEndPuncttrue
\mciteSetBstMidEndSepPunct{\mcitedefaultmidpunct}
{\mcitedefaultendpunct}{\mcitedefaultseppunct}\relax
\EndOfBibitem
\bibitem[te~Velde \latin{et~al.}(2001)te~Velde, Bickelhaupt, Baerends,
  Fonseca~Guerra, van Gisbergen, Snijders, and Ziegler]{ADF2001}
te~Velde,~G.; Bickelhaupt,~F.~M.; Baerends,~E.~J.; Fonseca~Guerra,~C.; van
  Gisbergen,~S. J.~A.; Snijders,~J.~G.; Ziegler,~T. Chemistry with ADF.
  \emph{J. Comput. Chem.} \textbf{2001}, \emph{22}, 931--967\relax
\mciteBstWouldAddEndPuncttrue
\mciteSetBstMidEndSepPunct{\mcitedefaultmidpunct}
{\mcitedefaultendpunct}{\mcitedefaultseppunct}\relax
\EndOfBibitem
\end{mcitethebibliography}

\end{document}